\documentclass{aa}  
\usepackage{graphicx,xcolor}
\usepackage{longtable,lscape}
\usepackage{dcolumn}
\usepackage{txfonts}
\usepackage[authoryear]{natbib} 
\bibpunct{(}{)}{,}{a}{}{,} 
\def\changed{\color{black}}
\begin{document}
   \title{The Quintuplet Cluster~\thanks{Based on observations
   collected at the ESO VLT ({\changed program} 077.D-0281(A)).}}

   \subtitle{I. A $K$-band spectral catalog of stellar sources} 

   \author{A. Liermann
          \and
          W.-R. Hamann
          \and
          L. M. Oskinova
          }


   \institute{University of Potsdam, Institute for Physics and Astronomy,
              14476 Potsdam, Germany
             }

   \date{Received June 11, 2008; accepted March 16, 1997}

   \titlerunning{The Quintuplet Cluster. I.}

\abstract
{
Three very massive clusters are known to reside in the Galactic Center
region, the Arches cluster, the Quintuplet cluster and the Central
Parsec cluster, each of them being rich in young hot stars. With new
infrared instruments this region is no longer obscured to the 
observer.
}
{For understanding these very massive clusters, it is essential to
know their stellar inventory. We provide comprehensive
spectroscopic data for the stellar population of the Quintuplet cluster
that will form the basis of subsequent spectral analyses.
}
{
Spectroscopic observations of the Quintuplet cluster have been
obtained with the Integral Field Spectrograph SINFONI-SPIFFI at the
ESO-VLT. The inner part of the Quintuplet cluster was covered by 22 slightly
overlapping fields, each of them of $8\,\arcsec \times 8\,\arcsec$ in size.
The spectral range comprises the near-IR $K$-band from 1.94 to
2.45\,$\mu$m. The 3D data cubes of the individual fields were
flux-calibrated and combined to one contiguous cube, from which the
spectra of all detectable point sources were extracted.
}
{
We present a catalog of 160 stellar sources in the
inner part of the Quintuplet cluster. The flux-calibrated $K$-band
spectra of 98 early-type stars and 62 late-type stars are provided as Online
Material. Based on these spectra, we assign spectral types to all
detected sources and derive synthetic $K_\textrm s$-band
magnitudes. Our sample is complete to about the 13th $K$-magnitude. We
report the detection of two hitherto unknown Wolf-Rayet stars of late
WC type (WC9 or later). Radial velocities are measured and employed to
asses the cluster membership. The quantitative analysis of the
early-type spectra will be subject to a subsequent paper.
}
{}

   \maketitle
%

\section{Introduction}
The Galactic Center has long been hidden from optical observation due to high
extinction in the visual. {\changed Today} the advanced instruments
for infrared radiation give access to this region.
Three surprisingly young and massive stellar clusters were
found within {\changed a projected distance of} 35\,pc from the
central black hole: 
the Arches, the Quintuplet and the Central cluster.  
These clusters are found to be rich in massive stars
(Arches:\,\citealt{Blum+2001, Figer2004, Martins+2008};
Quintuplet:\,\citealt{FigerMcLeanNajarro1997, FMM99, 
 Figer2004}; Galactic Center Cluster:\,\citealt{Eckart+2004, Figer2004}).
The prominent constellation of five infrared-bright stars gave
reason to name one of the {\changed clusters} the ``Quintuplet
cluster'' \,\citep{Okuda+1987, Okuda+1989}. It has a projected
distance of 30\,pc to the Galactic Center, a cluster radius of about
1\,pc, and an estimated age of about 4 million
years\,\citep{Okuda+1990, FMM99}.  

First spectroscopic observations of the five ``Quintuplet proper''
stars {\changed \citep[later named Q1, Q2, Q3, Q4 and Q9 by][]{FMM99}}
showed only a continuum without any features.  
Some of them have been detected also as far-IR and radio sources
\citep{Lang+1997, Lang+1999, Lang+2005}.  \citet{Tuthill+2006}
resolved two of the original Quintuplet stars (Q2 and Q3) in space
and time as colliding wind binaries, ejecting dust in the shape of a
rotating ``pinwheel''.  According to further spectroscopic studies
\citep{FMM99, Figer2004}, the cluster contains at least 100 O stars
and 16 Wolf-Rayet stars.  Two massive{\changed ,} evolved stars in
their luminous blue variable {\changed (LBV)} phase, the ``Pistol
star'' and one further LBV candidate, have been found in the
Quintuplet cluster \citep{Geballe+2000}.  {\changed At least two
luminous} stars of late spectral type K and M {\changed were} found in the cluster as well.

The present paper {\changed (hereafter the ``LHO catalog'')} provides a comprehensive
spectral atlas of $K$-band spectra from stellar sources in the
Quintuplet cluster. These spectra can form the basis for a subsequent
analysis of the luminous stellar population, which is {\changed a}
prerequisite for understanding the formation and evolution of this
very massive cluster in its special galactic environment.

The paper is structured as follows: Section\,\ref{sec:observations}
gives details on the observations and data
reduction. Section\,\ref{sec:catalog} contains the catalog with 
position, spectral classification, synthetic
K magnitude and radial velocity. The final Sect.\,\ref{sec:summary}
summarizes the catalog. The spectral atlas
comprising 160 stars is available as Online Material.

\section{Observations}
\label{sec:observations}
\subsection{The data}
We obtained Service Mode observations with
the ESO VLT UT4 (Yepun) {\changed telescope} between May and July 2006
using the integral field spectrograph SPIFFI of the SINFONI module that is
equipped with a Rockwell Hawaii 2RD 2k$\times$2k detector
\citep{Eisenhauer+2003, Bonnet+2004}. 
The Adaptive Optics facility could not be used since no suitable guide
star is available in this field and the laser guide star was not
offered {\changed during} this period. 

   \begin{figure*}[!ht]
   \centering
   \includegraphics[width=10.cm]{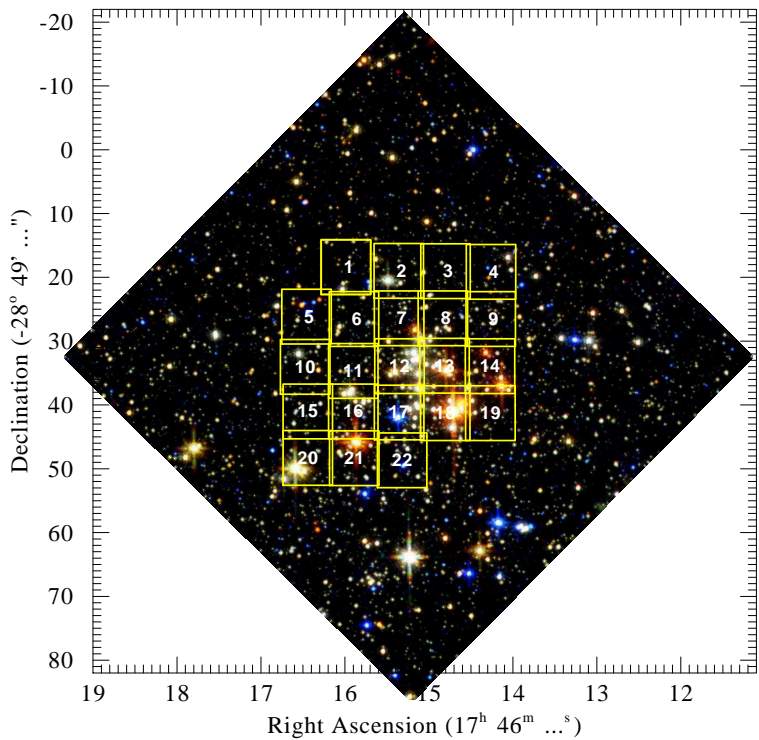}
   \caption{The 22 observed fields, overlayed on an HST composite
   image (HST Heritage archive)}
   \label{fig:HST+OBs}
   \end{figure*}

The observations consist of 22 observation blocks (OBs) covering the dense
center of the Quintuplet cluster with slightly overlapping fields of
view of 8\arcsec $\times$ 8\arcsec (see Fig.\,\ref{fig:HST+OBs}). ABBA
cycles were performed with an 
exposure time of 2 $\times$ 5\,min on each science field. The
grating for the $K$-band (1.95 - 2.45\,$\mu$m) provides a
spectral resolution of $R \approx 4000$. Table\,\ref{tab:list-of-OBs}
gives an overview {\changed of} the observations with their date and the
center coordinates of the FoV.
In the same mode, standard stars were observed at similar airmass 
for the flux calibration of the spectra.

\subsection{Data reduction}
The first steps of the data reduction were performed using the SPRED
software \citep{Abuter+2006}.
Raw-frames are corrected for bad pixels, distortion and
flatfield, followed by the wavelength calibration. The slitlets
are stacked into a data cube, which then consists of
layers of monochromatic images.
This reduction process was applied to the science exposures as well
as to the standard star observations. 

   \begin{table}
      \caption[]{List of observation blocks (OBs)}
      \label{tab:list-of-OBs}
      \begin{center}
         \begin{tabular}{llccc}
            \hline \hline
            \noalign{\smallskip}
            OB  & Name of OB & Date & R.A. & Dec.  \\
            && 2006& 17$^\mathrm h$ 46$^\mathrm m$ & -28$^\circ$ 49$\arcmin$ \\
            \noalign{\smallskip}
            \hline
            \noalign{\smallskip}
            229060 & Quintuplet-01 & {\changed 18-05} & 15\fs99 & 18\farcs4 \\
            229062 & Quintuplet-02 & {\changed 01-06} & 15\fs36 & 19\farcs0 \\
            229064 & Quintuplet-03 & {\changed 01-06} & 14\fs80 & 19\farcs0 \\
            229066 & Quintuplet-04 & {\changed 07-06} & 14\fs26 & 19\farcs1 \\
            229068 & Quintuplet-05 & {\changed 07-06} & 16\fs46 & 26\farcs1 \\
            229070 & Quintuplet-06 & {\changed 07-06} & 15\fs89 & 26\farcs5 \\
            229072 & Quintuplet-07 & {\changed 24-06} & 15\fs35 & 26\farcs4 \\
            229074 & Quintuplet-08 & {\changed 24-06} & 14\fs83 & 26\farcs4 \\
            229076 & Quintuplet-09 & {\changed 24-06} & 14\fs26 & 26\farcs5 \\
            229078 & Quintuplet-10 & {\changed 24-06} & 16\fs47 & 34\farcs0 \\
            229080 & Quintuplet-11 & {\changed 24-06} & 15\fs90 & 34\farcs7 \\
            229082 & Quintuplet-12 & {\changed 24-06} & 15\fs37 & 33\farcs9 \\
            229084 & Quintuplet-13 & {\changed 24-06} & 14\fs81 & 33\farcs9 \\
            229086 & Quintuplet-14 & {\changed 24-06} & 14\fs27 & 33\farcs9 \\
            229088 & Quintuplet-15 & {\changed 24-06} & 16\fs44 & 41\farcs0 \\
            229090 & Quintuplet-16 & {\changed 24-06} & 15\fs90 & 41\farcs0 \\
            229092 & Quintuplet-17 & {\changed 29-06} & 15\fs35 & 41\farcs2 \\
            229094 & Quintuplet-18 & {\changed 29-06} & 14\fs81 & 41\farcs3 \\
            229096 & Quintuplet-19 & {\changed 29-06} & 14\fs27 & 41\farcs3 \\
            229098 & Quintuplet-20 & {\changed 30-06} & 16\fs44 & 48\farcs3 \\
            229100 & Quintuplet-21 & {\changed 01-07} & 15\fs89 & 48\farcs3 \\
            229102 & Quintuplet-22 & {\changed 01-07} & 15\fs32 & 48\farcs7 \\
            \noalign{\smallskip}
            \hline
         \end{tabular}
      \end{center}
   \end{table}

\begin{figure*}
   \centering
   \includegraphics[width=\columnwidth, angle=-90]{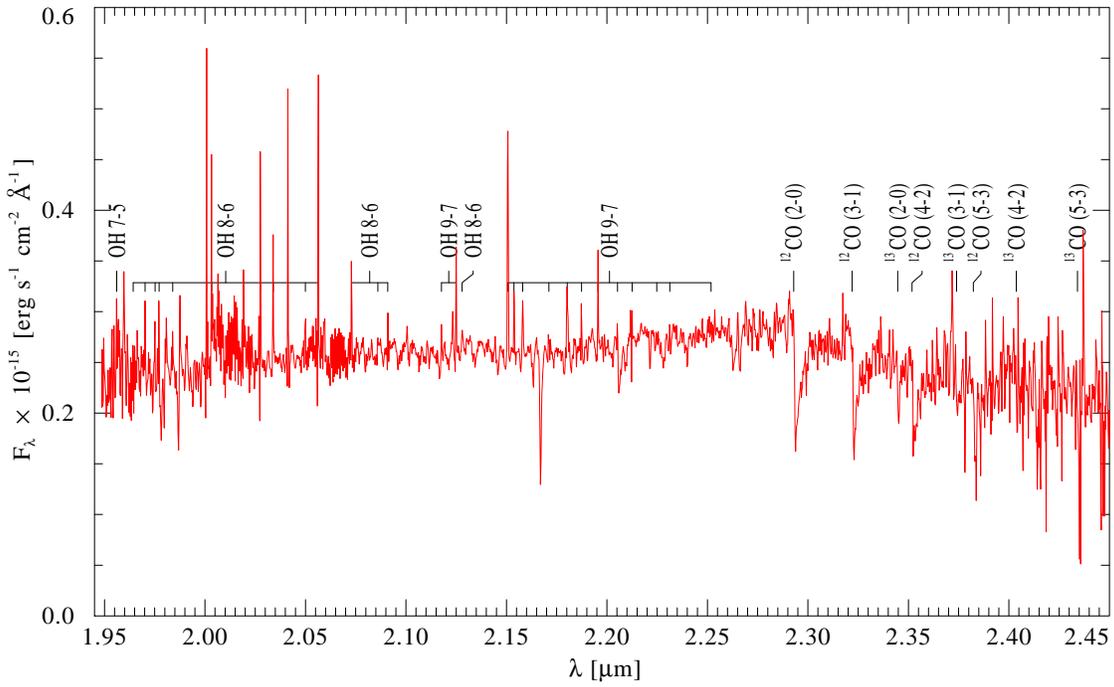}
      \caption{{\changed Flux-calibrated} $K$-band spectrum of an M star
      (LHO\,4). Identifications for the telluric OH lines are from
      \citet{Rousselot+2000}. The prominent molecular absorption bands
      are produced by CO. Their identifications are based on
      \citet{Gorlova+2006} and \citet{Wallace-Hinkle1997}. Note the
      discrimination between the isotopes $^{12}$CO and $^{13}$CO.}
         \label{fig:OH+Mstar}
   \end{figure*}

Sky subtraction was not performed at this stage of the data reduction,
since the sky frames turned out to be contaminated 
with stellar sources that would create ``negative stars'' in the
reduced data. {\changed Therefore sky frames could not be applied}. 
The sharp emission peaks in the extracted
spectra, especially of the fainter objects, are due to {\changed
telluric OH} contamination (see Fig.\,\ref{fig:OH+Mstar}){\changed .}

Further data reduction was performed with a self-written
{\sc{IDL}} cascade consisting of three parts.
First the standard star data cubes were sky-subtracted, using 
for each wavelength layer the monochromatic median of a pre-defined ``star 
free'' sub-array of the field. 
Then, the standard star spectra were extracted by adding all pixels
within a predefined extraction radius, mimic{\changed k}ing a synthetic
ape{\changed r}ture. {\changed Each standard star spectrum was identified with a
Kurucz model \citep{Kurucz1993}}, selected according 
to the spectral type (between B and G) and scaled to the
2MASS $K_\textrm s$-band magnitude \citep{2MASS}. {\changed The
calibration curve, that will be applied later to the science
observations, was obtained as the ratio between the model flux and
the extracted standard-star spectrum.} As the Kurucz models do not perfectly 
reproduce the Br$\gamma$ line profile, the residual feature in
{\changed the} 

   \begin{figure}[!ht]
   \centering
   \includegraphics[width=7.cm]{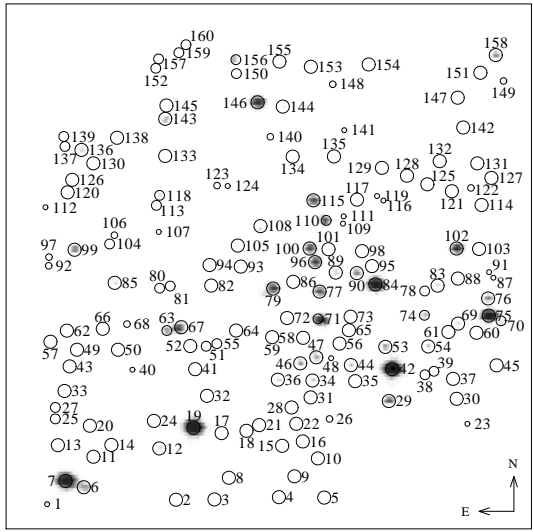}
   \caption{Image of the collapsed Grand Mosaic Cube. Open circles
            mark the detected sources with their running number in
            this ``LHO'' catalog. Circle radii correspond to the
            extraction radius for the object spectra.}  
              \label{fig:map}
   \end{figure}
   \begin{figure}[!ht]
   \centering
   \includegraphics[width=8.cm]{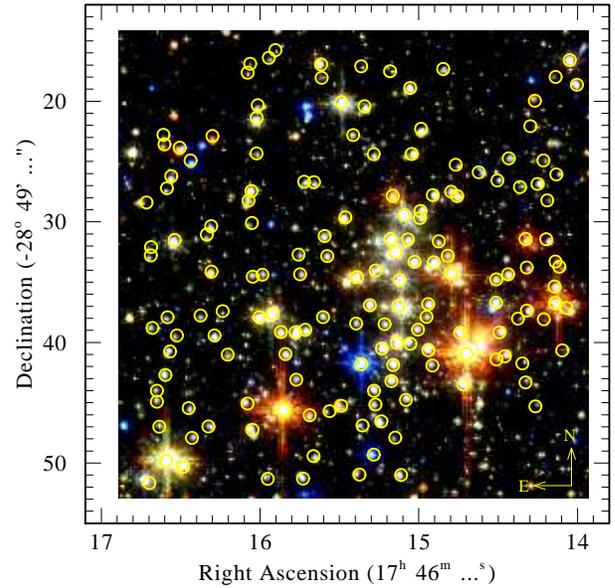}
      \caption{Detected sources in the Quintuplet cluster as overlay
      on a detail of the HST composite image (compare Fig.\,\ref{fig:map}
      for object identification{\changed, figures not to scale}).}  
         \label{fig:HST+stars}
   \end{figure}
calibration curve was smoothed out by interpolation
between 2.15 and 2.17\,$\mu$m. 

{\changed Sky subtraction for the science observations was performed as well 
with a monochromatic median of a pre-defined ``star-free'' subarray
of each cube}. Subsequently, each {\changed science observation} was
flux-calibrated by {\changed multiplying the cube wavelength plane
with the corresponding value of the calibration curve.
Then, all OBs were combined to one mosaic.
The nominal telescope pointing turned out to be too imprecise
for this purpose. Instead, we
generated 2D images for each OB by ``collapsing'' the data 
cube to a passband image. These images were then used to determine the
mutual spatial offsets between the fields by manually overlaying them
on the HST composite image by D. Figer (HST program 7364, STScI-PRC1990-30b).}
Applying these offsets, all 22 {\changed OB} cubes were combined into one
Grand Mosaic Cube, averaging the flux for pixels in overlapping regions. 

The final part of the IDL cascade takes this Grand Mosaic Cube 
{\changed and collapses it to a passband image. Then an
automated search for detecting point sources was run on this 2D image.
The radius to extract point sources was generally set to 4 pixels. A
smaller synthetic aperture was chosen for stars that have close
neighbors or are at the borders of the Grand Mosaic Cube.}
In any case the applied calibration curve was based on 
standard star spectra extracted with the same aperture.
Figure\,\ref{fig:map} shows the {\changed passband}
image of the Grand Mosaic Cube overlayed with the detected sources and
their running number in this ``LHO'' catalog.
   \begin{figure*}[!ht]
   \centering
   \includegraphics[width=\columnwidth, angle=-90]{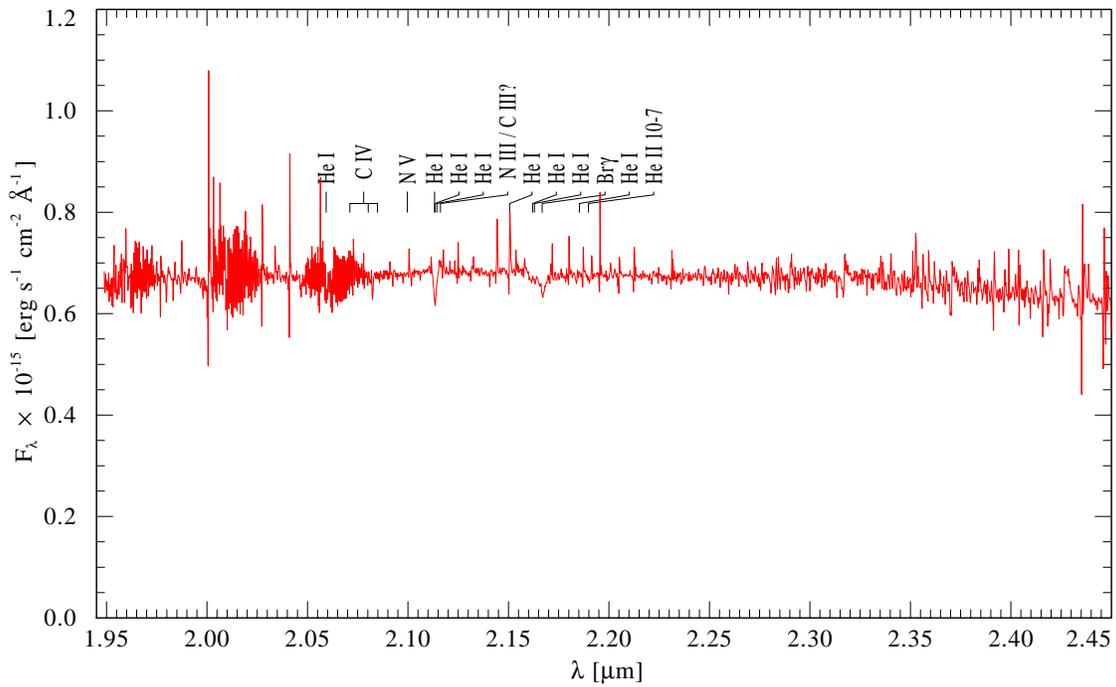}
      \caption{{\changed Flux-calibrated} $K$-band spectrum of an O
      star (LHO\,55) with line identifications as used for spectral
      classification.}
         \label{fig:Ostar}
   \end{figure*}
   \begin{figure*}[!ht]
   \centering
   \includegraphics[width=\columnwidth, angle=-90]{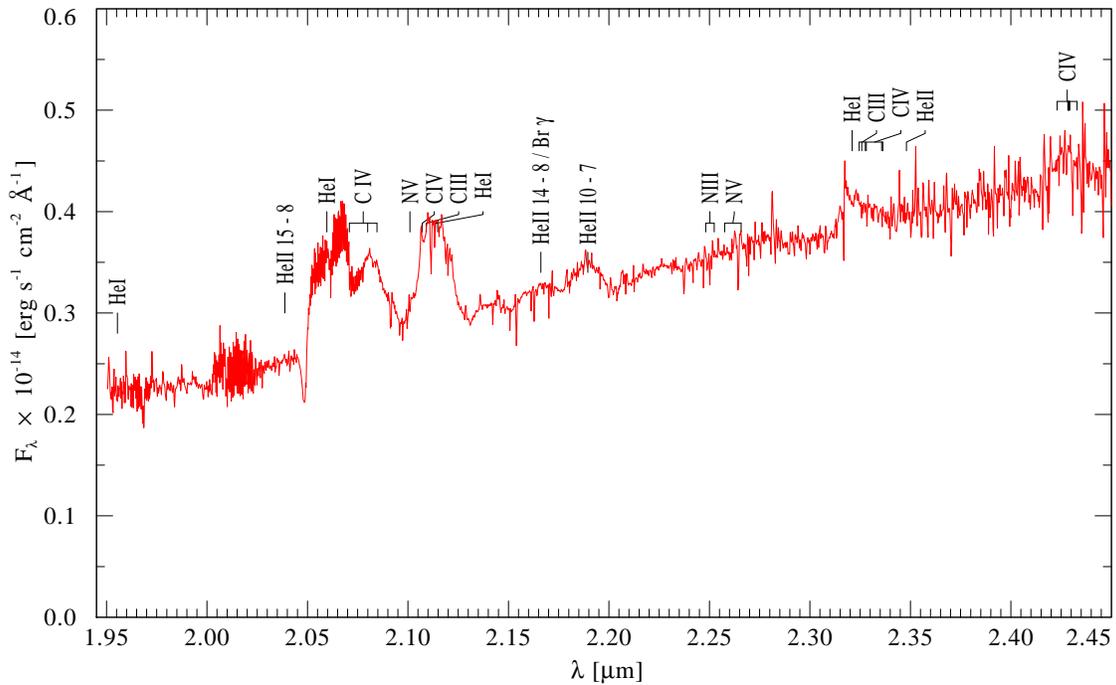}
      \caption{{\changed Flux-calibrated} $K$-band spectrum of the
      newly detected WR star LHO\,76 with line identifications. The
      spectral type is WC9.}
         \label{fig:newWRstar}
   \end{figure*}

   \begin{figure*}
   \centering
   \includegraphics[width=\columnwidth, angle=-90]{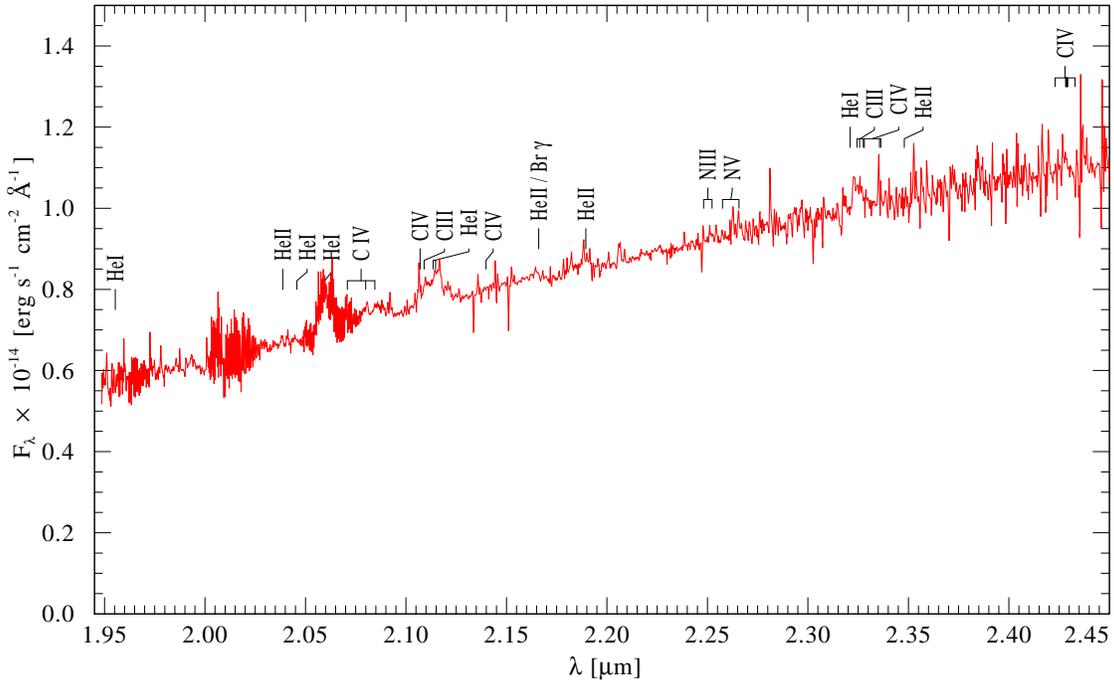}
      \caption{{\changed Flux-calibrated} $K$-band spectrum of Q\,6 (LHO\,79) showing broad
      emission line features superimposed on a warm-dust
      continuum. The spectral class is WC9, revising the
      classification by \citep{FMM99} who gave ``$<$ B0 I'' for the
      alias-named object qF\,250.} 
         \label{fig:QPMspectrum}
   \end{figure*}

\section{The catalog}
\label{sec:catalog}
\subsection{Coordinates and identification}

{\changed In order to derive} the coordinates, we superimpose the map
of the detected sources (Fig.\,\ref{fig:map}) on the HST image of the field
(Fig.\,\ref{fig:HST+OBs}). The well determined positions of foreground
stars from the USNO-B catalog \citep{USNO} are used to construct the
mapping between the pixel positions and the celestial coordinates.
An overlay image with coordinates is shown in Fig.\,\ref{fig:HST+stars}. 

Table\,\ref{tab:catalog} lists all {\changed 160} extracted sources.
Column 1 gives the running number in this ``LHO'' catalog, followed by
the coordinates in column 2 and 3. The SIMBAD database was employed to
find possible cross-identifications (see column 7).

{\changed \citet{Lang+2005} reported that for their radio sources
QR~1, QR~2 and QR~3 no counterparts are found in the near-IR. This led
these authors to discuss the possibility that the sources are
ultracompact \ion{H}{ii} regions with ongoing star formation. 
We confirm the absence of K-band counterparts for these radio sources
in our data.}

\subsection{Spectral classification}
\label{subsec:spec-class}
Classification of OB stars was performed according to \citet{Hanson+1996,
Hanson+2005} and \citet{Morris+1996}, concentrating on the most
prominent lines in the $K$-band region (\ion{H}{i} 2.1661\,$\mu$m
Br$\gamma$, \ion{He}{i} 2.0587\,$\mu$m, \ion{He}{ii}
2.1891\,$\mu$m, \ion{C}{iv} triplet around 2.0796\,$\mu$m,
\ion{N}{iii}/\ion{C}{iii} 2.1155\,$\mu$m) for comparison with their
template spectra. Additional \ion{He}{i} lines (2.1127/37\,$\mu$m,
2.1499\,$\mu$m, 2.1623\,$\mu$m) were employed to distinguish
supergiants from giants. Spectral peculiarities are given in
Table\,\ref{tab:catalog} in the classical nomenclature and are
followed by ``?'' in uncertain cases such as noisy spectra or blends
with atmospheric OH lines.  
An example of an O star spectrum is shown in Fig.\,\ref{fig:Ostar}.  

Wolf-Rayet stars have been classified following the criteria from
\citet{Crowther+2006}. For WC stars we measured line equivalent widths
and determined the ratio \ion{C}{iv} 2.079\,$\mu$m/\ion{C}{iii}
2.108\,$\mu$m. 
The ratio \ion{He}{ii}\,(2.189\,$\mu$m)/Br$\gamma$\,(2.166\,$\mu$m) was
used for the subtype classification of WN stars.
A strongly increasing continuum with wavelength was considered to
indicate the dominance of warm dust emission, marked with ``d'' in the
spectral classification.

Our sample contains eleven previously known WR stars, four WN and seven
WC. Two new WR stars are found in our sample, LHO\,76 and 79.

We could not identify LHO\,76 with any previously known
source. Therefore we claim that this object is a newly discovered WR
star in the Quintuplet cluster; its spectral type is WC9d
(Fig.\,\ref{fig:newWRstar}). Following the naming scheme of 
\citet{vdH2001}, this star would {\changed squeeze in between the
known stars WR\,102da and WR\,102db. A possible scheme to rename this
group could be LHO\,76 $\rightarrow$ WR\,102db (new), WR\,102db (old)
$\rightarrow$ WR\,102dc (new), WR\,102dc (old) $\rightarrow$ WR\,102dd
(new), WR\,102dd (old) $\rightarrow$ WR\,102de (new).} 

The other {\changed new WR} star - LHO\,79 - was already listed by
\citet{FMM99} as qF\,250 but classified as $<$ B0 I. We re-classify
this star as spectral type WC9d (see
Fig.\,\ref{fig:QPMspectrum}). {\changed This increases the total
number of WR stars in the sample to 13}. 

All WC stars turned out to be of late subtype, WC8 and WC9. In all
cases the WC9 spectra {\changed display} dust emission, while the two WC8
spectra (LHO\,34 and 47) are free of such contamination. 

The stellar spectra of LHO\,75 and 102 are {\changed particularly}
drowned in dust emission. The diluted stellar lines appear unusually
narrow. The identification of the classification line \ion{C}{iii}
2.110\,$\mu$m is questionable and therefore the WC9 subtype is
uncertain (``WC9?d'') for these two stars. 

{\changed Late-type} stars were classified using the dominating features of
the absorption bands caused by the CO first overtone. We find both 
$^{12}$CO and $^{13}$CO absorption, the later indicating late
type (super-)giant stars \citep{Goorvitch1994, Wallace-Hinkle1997,
Kleinmann-Hall1986}. Fig.\,\ref{fig:OH+Mstar} shows LHO\,4 as an
example {\changed of} an M star.
We then measured the equivalent width of the band-head CO (2-0) to
determine the effective temperature 
\begin{center}
\begin{equation}
T_\mathrm{eff} = 4895 - 62 \times \textrm{EW(CO)}
\end{equation}
\end{center}
and spectral subtypes of the stars 
\begin{center}
\begin{equation}
G = 0.56 \times \textrm{EW(CO)} - 3.0 \,,
\end{equation}
\end{center}
following \citet{Gonzalez-Fernandez+2008}. The integer number $G$
ranges from 0 to 13 representing the subtypes from K0 to M7, e.g. $G
=6$ gives M0. Derived $G$ values were rounded to the next integer to
{\changed determine the} subtypes {\changed as listed in
Table\,\ref{tab:catalog}}.  Both relations are only applicable for
(super-)giant stars, an assumption that {\changed is} justified by the
{\changed presence} of the $^{13}$CO band absorption mentioned above{\changed,}
as well as by the synthetic $K_\textrm{s}$ magnitudes, see
Sect.\,\ref{subsec:photometry}.  For seven stars the measured
equivalent width was so small that the resulting $G$ value was
slightly in the negative range. We classify these stars as ``K0?'' in
Table\,\ref{tab:catalog} indicating that they might be of earlier
spectral subtype.

\subsection{Photometry}
\label{subsec:photometry}
\begin{figure}[t]
   \centering
   \includegraphics[width=6.cm, angle=-90]{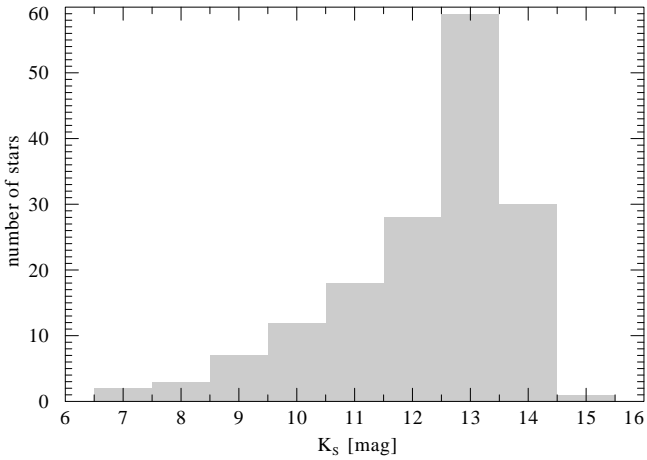}
      \caption{Histogram of the synthetic $K$-band magnitudes
      {\changed with a bin size of 1\,mag},
      indicating a photometric completeness of the catalog up to
      $K_\textrm s$ = 13\,mag.}
         \label{fig:histogramm}
   \end{figure}

From our flux calibrated spectra we derive synthetic $K$-band magnitudes.
For comparability we choose the $K_\textrm s$ ($K$ ``short'') filter of
2MASS which is centered at 2.159\,$\mu$m, and use the calibration
constant from \citet{Cohen+2003} for the photometric zero
point. {\changed Fifteen} LHO stars are in common with the sample of
\citet{GMC99}. We compare our derived magnitudes with their values and
obtain {\changed mean} agreement within {\changed 0.4\,mag}.
Obtained magnitudes for all catalog objects are listed in
Table\,\ref{tab:catalog}. The number of stars per magnitudes is shown
in Fig.\,\ref{fig:histogramm} as histogram. {\changed The distribution
follows a power-law up to $K_\textrm s$ = 13\,mag, 
which indicates that our catalog is complete to this magnitude.}

   \begin{figure}[!hb]
   \centering
   \includegraphics[width=\columnwidth]{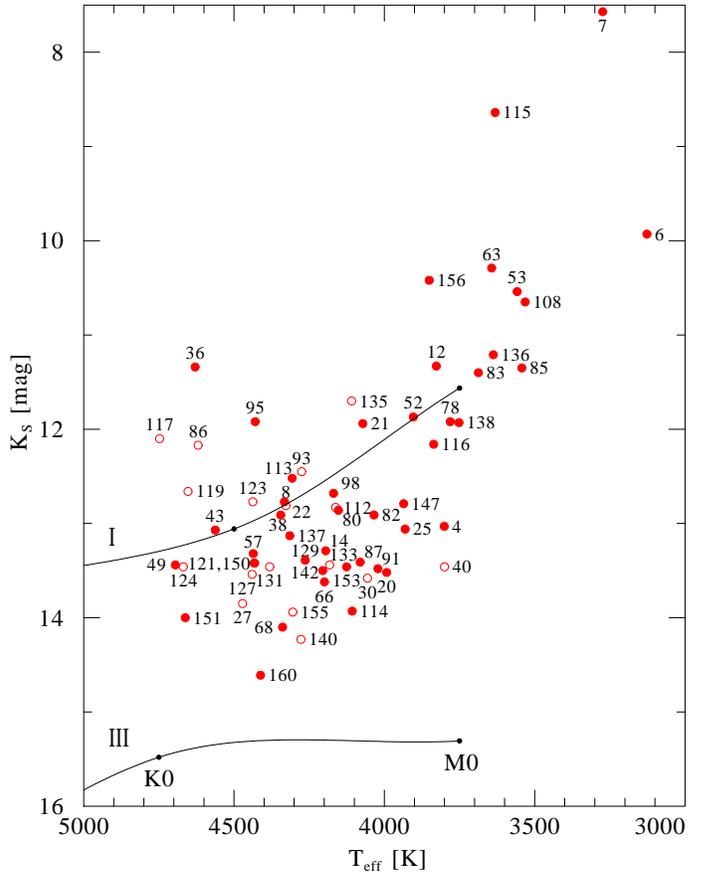}
      \caption{$K$ magnitude vs. $T_\mathrm{eff}$ for the {\changed late-type}
      stars of our sample, identified by their LHO number
      (Table\,\ref{tab:catalog}). Open circles denote stars for which  
      the radial velocity is uncertain or definitely indicates that
      they are foreground objects. 
      The solid line interpolates between corresponding values from
      Kurucz models for G0, K0 and M0 stars. The upper line refers to
      supergiant models (luminosity class I) and the lower line to
      giants (luminosity class III).} 
      \label{fig:COgiants}
   \end{figure}

{\changed To determine} the approximate luminosity class of our
{\changed late-type}
stars we plotted their $K_\textrm{s}$ magnitudes {\changed versus}
their determined effective temperature (Fig.\,\ref{fig:COgiants}). For
comparison we take models \citep{Kurucz1993} of G0, K0 and M0 stars of the
luminosity classes I and III, with assumed masses of
1.1\,M$_\odot$, 0.8\,M$_\odot$ and 0.4\,M$_\odot$, respectively. The
model fluxes are scaled for the adopted cluster distance of 8\,kpc
\citep{Reid1993} and diminished according to the average extinction
of $A_K = 3.28$\,mag \citep{FMM99}. The models are connected in
Fig.\,\ref{fig:COgiants} by solid lines to mark the approximate
regions of the corresponding luminosity class. 
Most of our sample stars scatter around the line for supergiants. Some
stars are slightly less bright and {\changed therefore} are assigned
to luminosity class II. A few stars{\changed , LHO\,6, LHO\,7 and
LHO\,115,} seem to be extremely bright{\changed, see Fig.\,\ref{fig:COgiants}}. 

\subsection{Cluster membership}
\label{subsec:clustermembers}
\begin{figure}[!ht]
   \centering
   \includegraphics[width=8.cm]{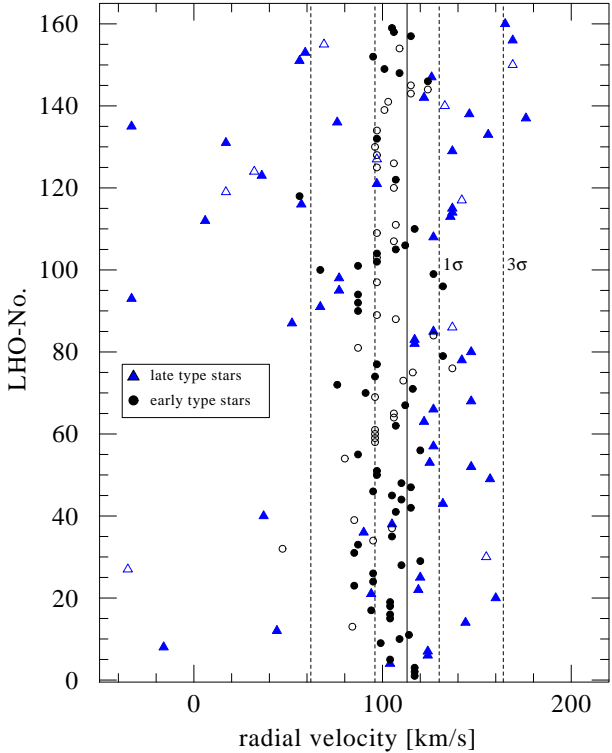}
      \caption{Radial velocities of the catalog stars. Circles denote
      {\changed early-type} stars, triangles {\changed late-type}
      stars. Open symbols are for uncertain values due to noisy
      spectra. The solid vertical line indicates {\changed
      $\overline{RV}_\mathrm{QPM} = 113$\,km/s, while the dashed lines
      indicate intervals of $\pm\sigma$ and $\pm 3\sigma$ ($\pm
      17$\,km/s and $\pm 51$\,km/s), respectively.}} 
         \label{fig:RV}
   \end{figure}

For all spectra in the catalog we measured the wavelength of a
prominent feature, Br\,$\gamma$ for {\changed early-type} stars and CO
(2-0) for {\changed late-type} stars, and derived the radial velocity
for the individual objects (Fig.\,\ref{fig:RV} and
Table\,\ref{tab:catalog}). {\changed Heliocentric corrections have
been applied to the radial velocities.}  The {\changed  
measurement errors} can be expected to {\changed be maximal} $\pm
37$\,km/s {\changed from} the nominal spectral resolution of the
instrument.  The results may be more uncertain for some of the more
noisy spectra (marked with ``:'' in Table\,\ref{tab:catalog}).

Assuming that the {\changed fifteen bright} Quintuplet stars from the
QPM list \citep{GMM90} are real cluster members, we derive their mean
radial velocity {\changed $\overline{RV}_\mathrm{QPM} =
113$, with a standard deviation of $\sigma = 17$\,km/s}. This is in
good agreement with the 130\,km/s found by \citet{FMM99}.  
{\changed For a prediction of the internal velocity dispersion of the
cluster, we follow \citet{Kroupa2002}. Despite the reported young
age, we assume a virialized cluster, a star formation efficiency between
20 and 40\,\%  \citep{Kroupa2002}, and a stellar mass in the range of
$10^{3.8-4.2}\,\mathrm{M}_\odot$ \citep{FMM99}.
This yields a radial velocity dispersion of 8-18\,km/s, which agrees well with the determined standard deviation
of the fifteen QPM stars, $\sigma = 17$\,km/s.} 
Therefore we consider {\changed the $3\sigma$ interval around
$\overline{RV}_\mathrm{QPM}$} as a {\changed criterion for} cluster
membership. {\changed As Fig.\,\ref{fig:RV} shows, there are a few} stars in 
our sample with smaller radial velocities, {\changed primarily} of
late spectral type. They {\changed have to be} considered as
foreground objects, i.e.\ K or M dwarfs, and are marked ``f'' in
Table\,\ref{tab:catalog}. 

\section{Summary}
\label{sec:summary}
We present {\changed the
first} $K$-band spectral catalog of {\changed 160} stellar sources in
the central region of the Quintuplet cluster. Integral field {\changed
spectra} were
obtained with the SINFONI-SPIFFI on UT4 from May to July 2006. Our
catalog comprises {\changed flux-calibrated spectra of} 98 {\changed
early-type} stars (OB and WR) and 62 {\changed late-type} stars (K to
M) {\changed with a spectral resolution of $R \approx 4000$.}
Table\,\ref{tab:number-statistics} lists the number of stars per
spectral class. 

   \begin{table}[!ht]
      \caption[]{Spectral type distribution}
      \label{tab:number-statistics}
      \begin{center}
         \begin{tabular}{lr}
            \hline \hline
            \noalign{\smallskip}
            Spectral Type & No. of stars \\
            \noalign{\smallskip}
            \hline
            \noalign{\smallskip}
            WN             & 4      \\
            WC             & 9  \\ 
            {\it{WR total}}& {\it{13}} \\ \hline
            \noalign{\smallskip}
            O              & 60 \\
            B              & 25 \\ 
            {\it{OB total}}& {\it{85}} \\ \hline
            \noalign{\smallskip}
            K              & 43 \\
            M              & 19  \\ 
            {\it{KM total}}& {\it{62}} \\ \hline
            {\it{total}}   & 160 \\
            \hline
         \end{tabular}
      \end{center}
   \end{table} 
%
All {\changed 160} sources are listed in Table\,\ref{tab:catalog} with their
{\changed LHO} number that was assigned to the source in our data reduction and
identification processes (column 1), the determined coordinates of the
source ({\changed columns} 2 and 3) and the derived synthetic $K_\textrm s$
magnitude (column 4). The spectral type is given in column 5. 
{\changed 
Radial velocities (RV) are given in column 6, together with an assessment
if this RV value is indicating a foreground object (``f''). Finally we
give a complete list of 
alias names and cross-identification with previous catalogs and
surveys in colum 7. 
Note that about 100 objects of our list are cataloged here for the
first time.} 

Two new WR stars {\changed of WC spectral subtype} are found in our sample,
LHO\,76 and 79, increasing the number ratio WC:WN in the Quintuplet
cluster to {\changed 10:6} in total (9:4 contained in our field). WC stars are
outstandingly frequent and bright in the Quintuplet cluster. For
comparison, {\changed the WC:WN number ratios from \citet{vdH2006} are
0:17 for the Arches cluster, 12:17 for the Central cluster, and
8:19 for Westerlund~I, respectively. A} difference {\changed to} the solar
neighbourhood {\changed WC:WN} ratio of 38:25 {\changed
\citep{vdH2001} can be seen.} A detailed analysis of the WR star
sub-sample with the Potsdam models for expanding atmospheres (PoWR) is
in preparation. 

{\changed Among the new stars in our catalog there are a number of late-type
supergiants. In case that the red supergiants (RSG) are real
cluster members, as supported by their radial velocities, 
the question of age and coeval evolution of the Quintuplet cluster has to
be readdressed in this context.}

The spectral atlas of the catalog is only available as Online
Material. It comprises for each detected 
source the flux-calibrated $K$-band spectrum in the wavelength range
from 1.95 to 2.45\,$\mu$m. The spectra are binned to 4\,\AA\,.
Some spectra suffer from atmospheric OH emission lines. 

\begin{acknowledgements}
  {\changed We thank the anonymous referee for helpful suggestions on
  the manuscript.}
  {\changed Furthermore,} the authors thank A. Barniske for the
  valuable help in 
  preparing the Phase 2 material. T. Szeifert and D. N\"urnberger
  supported A.~L. with very useful discussions on the ESO
  SINFONI-SPIFFI instrument, data reduction pipelines, its results
  and technical details in handling the huge amount of data.  Further
  on A.~L. thanks H. Todt for supporting, discussing and writing IDL
  routines on Integral Field Spectroscopy data in the reduction
  process. This publication makes use of data products from the Two
  Micron All Sky Survey, which is a joint project of the University of
  Massachusetts and the Infrared Processing and Analysis
  Center/California Institute of Technology, funded by the National
  Aeronautics and Space Administration and the National Science
  Foundation. This research has made use of the SIMBAD database,
  operated at CDS, Strasbourg, France.  A. Liermann is supported by
  the Deutsche Forschungsgemeinschaft (DFG) under grant HA 1455/19.

\end{acknowledgements}

\bibliographystyle{aa}
\bibliography{library}

\def\s{\rule[0mm]{0mm}{4mm}}
{\small{
\longtabL{3}{
\begin{landscape}
\begin{longtable}{cccrlll}
\caption{\label{tab:catalog} Catalog of Stars detected in the central Quintuplet cluster field}\\
\hline \hline \s
  LHO &  R.A. 17$^\textrm h$ 46$^\textrm m$ & Dec. -28\,$^\circ$
  49\,$\arcmin$ & $K_\textrm s$ &Spectral type& RV & Alias names \& remarks \\
  No. &  [s]   & [$\arcsec$]  & [mag] &  & [km/s] &\\ \hline
\endfirsthead
\caption{Continued.}\\
\hline \hline \s
  LHO &  R.A. 17$^\textrm h$ 46$^\textrm m$ & Dec. -28\,$^\circ$
  49\,$\arcmin$ & $K_\textrm s$ &Spectral type& RV & Alias names \& remarks \\
  No. &  [s]   & [$\arcsec$]  & [mag] &  & [km/s]&\\ \hline
\endhead
\hline
\endfoot
\s
   1& 16.70  & 51.4   & 10.9 & O3-8 I fe& {\changed 117}&[FMG99] 8 em. line star\\
   2& 15.95  & 51.3   & 13.2 & O7.5-B2 I-II e& {\changed 117}&\\
   3& 15.73  & 51.3   & 12.5 & O6-8 I f& {\changed 117}&\\
   4& 15.37  & 51.0   & 13.0 & M1 II & {\changed 104}&\\
   5& 15.11  & 51.0   & 12.6 & O6.5-9 I f?& {\changed 104}&\\
   6& 16.49  & 50.3   &  9.9 & M7 I & {\changed 124}\\
   7& 16.59  & 49.8   &  7.6 & M6 I & {\changed 124}& {\bf{Q~7}}, qF~192, NSV 23780, MGM~5-7\\
   8& 15.66  & 49.4   & 12.8 & K2 I & {\changed \,-16}f&\\
   9& 15.28  & 49.3   & 13.6 & O4-6 I f?& {\changed ~$\:$99}&\\
  10& 15.15  & 47.9   & 14.0 & O7-9 I f & {\changed 109}&\\
\s     			 	 
  11& 16.42  & 47.9   & 13.4 & O3-7 I-II e& {\changed 114}&\\
  12& 16.05  & 47.2   & 11.3 & M1 I & {\changed ~$\:$44f} &[GMC99] D~3605\\
  13& 16.64  & 47.0   & 13.5 & O6-B2 I & {\changed ~$\:$84}:&\\
  14& 16.33  & 47.0   & 13.3 & K3 II & {\changed 144}&\\
  15& 15.35  & 46.9   & 13.8 & O4-7 I f& {\changed 104}&\\
  16& 15.24  & 46.5   & 11.9 & O8.5-9.7 I ab? f?& {\changed 104}& qF~197?\\ 
  17& 15.69  & 46.0   & 12.2 & O3-7 I-II f?& {\changed ~$\:$94}&\\ 
  18& 15.56  & 45.7   & 13.6 & O7-8 I f& {\changed 104}&\\
  19& 15.85  & 45.6   &  7.2 & WC8/9d +OB & {\changed 104}& {\bf{Q~3}}, {\changed GCS 4, MGM~5-3}, WR~102ha, qF~211, [LY2004] QX~3, {\changed
    [GSL2002] 66, [NHS93] 26} ``pinwheel'' star\\
  20& 16.45  & 45.5   & 13.5 & K5 II & {\changed 160}&\\
\s     			 	 
  21& 15.49  & 45.3   & 11.9 & K4 I & {\changed ~$\:$94}&\\
  22& 15.27  & 45.1   & 12.8 & K2 I & {\changed 119}& \\
  23& 14.27  & 45.3   & 13.6 & O3-6 I-II f& {\changed ~$\:$85}&\\
  24& 16.08  & 45.1   & 12.4 & O7-9 I f? e& {\changed ~$\:$95}&\\
  25& 16.65  & 44.9   & 13.1 & M0 II & {\changed 120}&\\
  26& 15.08  & 44.8   & 13.3 & O5-B0 I f?& {\changed ~$\:$95}&\\
  27& 16.65  & 44.0   & 13.9 & K1 II & {\changed \,-35}:f&\\
  28& 15.28  & 44.0   & 12.3 & O7-9 I e& {\changed 110}&\\
  29& 14.72  & 43.5   &  9.9 & O9-B2 I f? e& {\changed 120}&\\
  30& 14.33  & 43.3   & 13.6 & K5 II & {\changed 155}:&\\
\s     			 	 
  31& 15.17  & 43.2   & 12.1 & O9-B1 I f?& {\changed ~$\:$85}&\\ 
  32& 15.77  & 43.1   & 13.3 & B1.5-3 I f?& {\changed ~$\:$47}:f&\\
  33& 16.60  & 42.7   & 12.8 & O9-B3 I-II f& {\changed ~$\:$87}&\\
  34& 15.16  & 41.9   & 11.3 & WC8 & {\changed ~$\:$95}:& WR~102g, qF~235S\\
  35& 14.91  & 41.9   & 12.8 & O9-B2 I e?& {\changed 105}&\\
  36& 15.36  & 41.8   & 11.3 & K0? I & {\changed ~$\:$90}& USNO-B1.000611-0602187? \citep{USNO} \\
  37& 14.35  & 41.7   & 14.1 & O9-B3 I-II& {\changed 105}:&\\
  38& 14.51  & 41.4   & 12.9 & K2 II & {\changed 105}&\\
  39& 14.46  & 41.1   & 12.2 & O7-B1 I f?& {\changed ~$\:$85}:& [GMC99] D~3606\\
  40& 16.20  & 41.0   & 13.5 & M1 II & {\changed ~$\:$37}f&\\
\s     			 	 
  41& 15.84  & 41.0   & 12.5 & O9-B1 I f& {\changed 107}&\\
  42& 14.70  & 41.0   &  6.7 & WC9d + OB & {\changed 115}&{\bf{Q~2}}, WR~102dc,
  GCS~3-2, {\changed MGM~5-1}, qF~231, [LFG99] QR~7, [LY2004] QX~5, [NHS93] 24, ``pinwheel'' star\\ 
  43& 16.57  & 40.8   & 13.1 & K0 I & {\changed 132}&\\
  44& 14.94  & 40.6   & 11.1 & O7-9 I f?& {\changed 110}&\\
  45& 14.10  & 40.7   & 14.0 & O7-B1 I & {\changed 105}&\\
\s
  46& 15.23  & 40.5   & 10.7 & O7-B1 I f?& {\changed ~$\:$95}& {\changed MGM~5-11b?}\\
  47& 15.14  & 40.0   & 10.4 & WC8 & {\changed 115}& {\bf{Q~11}}, {\changed MGM~5-11a, [NWS90] G}, WR~102f, qF~235N, [FMG99] 2, possible binary \citep{vdH2006}\\
  48& 15.05  & 40.1   & 12.0 & O7.5-9.5 I f?& {\changed 110}&\\
  49& 16.52  & 39.4   & 13.4 & K0? I & {\changed 157}&\\
  50& 16.29  & 39.4   & 12.7 & O7-B1 I f& {\changed ~$\:$97}& [LY2004] QX~4?\\
\s     			 	 
  51& 15.78  & 39.2   & 11.4 & O7-9 I f& {\changed ~$\:$97}&\\
  52& 15.87  & 39.1   & 11.9 & M0 I & {\changed 147}&[GMC99] D~334\\
  53& 14.74  & 39.2   & 10.5 & M3 I & {\changed 125}&\\
  54& 14.49  & 39.2   & 11.6 & O7-9 I-II f?& {\changed ~$\:$80}:&\\
  55& 15.72  & 38.9   & 12.0 & O7-9 I f& {\changed ~$\:$87}&\\
  56& 15.00  & 38.9   & 12.6 & O7-9 I f?& {\changed 120}&\\
  57& 16.68  & 38.8   & 13.3 & K1 I & {\changed 127}&\\
  58& 15.22  & 38.5   & 12.9 & O9-B1 I-II p?& {\changed ~$\:$96}:&\\
  59& 15.39  & 38.5   & 13.3 & O9-B1 I & {\changed ~$\:$96}:&\\
  60& 14.21  & 38.1   & 13.0 & O9-B1 I f?& {\changed ~$\:$96}:&\\
\s     			 	 
  61& 14.38  & 38.0   & 13.6 & O7-9 I f?& {\changed ~$\:$96}:&\\
  62& 16.58  & 37.9   & 14.0 & O7-B1 I-II f& {\changed 107}&\\
  63& 16.00  & 37.9   & 10.3 & M2 I & {\changed 122}&\\
  64& 15.60  & 37.9   & 12.9 & O7-9 I f?& {\changed 106}:&\\
  65& 14.95  & 37.9   & 12.8 & O3-7 I eq?& {\changed 106}:&\\
  66& 16.38  & 37.8   & 13.6 & K3 II & {\changed 127}&\\
  67& 15.92  & 37.6   &  9.6 & WN9 & {\changed 112}& {\bf{Q~8}}, WR~102hb, MGM~5-8, qF~240\\
  68& 16.24  & 37.4   & 14.1 & K2 II & {\changed 147}&\\
  69& 14.32  & 37.4   & 11.5 & O6-9.7 I f?& {\changed ~$\:$96}:&\\
  70& 14.07  & 37.1   & 11.0 & O7-8 I f& {\changed ~$\:$91}&\\
\s     			 	 
  71& 15.13  & 37.0   &  8.8 & WN9 & {\changed 116}&{\bf{Q~10}}, {\changed MGM~5-10}, qF~241, WR~102ea, {\changed [NWS90] F}, [LFG99] QR 5\\
  72& 15.31  & 36.9   & 11.9 & O4-6 I eq?& {\changed ~$\:$76}&\\
  73& 14.94  & 36.8   & 11.7 & O6.5-7 I f?& {\changed 111}:&\\
  74& 14.51  & 36.7   & 10.9 & O9.5-B1 I ab? f& {\changed ~$\:$96}&[LY2004] QX~1, qF~242?, {\changed [GMC99] D~309}\\
  75& 14.14  & 36.7   &  7.9 & WC9?d & {\changed 116}:& {\bf{Q~1}}, WR~102da, qF~243, GCS~3-4, MGM~5-1, narrow and weak lines\\
  76& 14.15  & 35.4   & 10.3 & WC9d & {\changed 137}:& NEW!!!\\
  77& 15.12  & 34.8   &  9.6 & O6-8 I f eq& {\changed ~$\:$97}& {\bf{Q~12}}, MGM~5-12, qF~278, [NWS90]~E\\
  78& 14.51  & 34.8   & 11.9 & M1 I & {\changed 142}& qF~252?\\
  79& 15.39  & 34.6   &  9.3 & WC9d & {\changed 132}& {\bf{Q~6}}, qF~250, MGM~5-6, [NWS90]~B, re-classified\\
  80& 16.05  & 34.6   & 12.9 & K4 I-II & {\changed 147}&\\
\s     
  81& 15.99  & 34.4   & 13.2 & O3-4 II/III? f& {\changed ~$\:$87}:&\\
  82& 15.75  & 34.3   & 12.9 & K5 II & {\changed 117}&\\
  83& 14.44  & 34.4   & 11.4 & M2 I & {\changed 117}& qF~252?\\
  84& 14.80  & 34.2   &  7.8 & WC9d & {\changed 127}:& {\bf{Q~4}}, {\changed MGM~5-4}, WR~102dd, GCS~3-1, qF~251\\
  85& 16.31  & 34.1   & 11.4 & M3 I & {\changed 127}&\\
  86& 15.27  & 34.0   & 12.2 & K0? I & {\changed 137}:&\\
  87& 14.11  & 33.7   & 13.4 & K4 II & {\changed ~$\:$52f}&\\
  88& 14.32  & 33.8   & 13.3 & O3-4 III f& {\changed 107}:&\\
  89& 15.03  & 33.3   & 11.1 & O7.5-8.5 I f& {\changed ~$\:$97}& [LY2004] QX~2?\\
  90& 14.91  & 33.4   & 10.3 & O7-9.5 I f e?& {\changed ~$\:$87}& GCS~3{\changed I} \ion{H}{II} region?, [GSL2002]~63?\\

\s     
  91& 14.14  & 33.3   & 13.5 & K5 II & {\changed ~$\:$67}&\\
  92& 16.69  & 32.8   & 13.9 & O8-9.7 I & {\changed ~$\:$87}&\\
  93& 15.58  & 32.8   & 12.5 & K3 I & {\changed \,-33}f&\\
  94& 15.76  & 32.7   & 13.3 & O7-B1 I f?& {\changed ~$\:$87}&\\
  95& 14.82  & 32.8   & 11.9 & K1 I & {\changed ~$\:$77}& GCS~3{\changed I} \ion{H}{II} region?, [GSL2002]~63?\\
  96& 15.15  & 32.5   &  9.3 & O6-8 I f e& {\changed 132}& {\changed MGM~5-13b?}\\
  97& 16.69  & 32.1   & 14.1 & O7-B2 I-II & {\changed ~$\:$97}:&\\
  98& 14.87  & 31.6   & 12.7 & K4 I & {\changed ~$\:$77}&\\
  99& 16.54  & 31.5   & 10.1 & WN9 & {\changed 127}&WR~102i, qF~256, [GMC99] D~215  \\
 100& 15.18  & 31.4   &  9.4 & O6-8 I f e& {\changed ~$\:$67}&{\bf{Q~13}}, {\changed MGM~5-13a, [NWS90] D}, [LFG99] QR~6, q{\changed F}~257\\
\s     			 	 
 101& 15.07  & 31.5   & 11.9 & O6-8 I f?e?& {\changed ~$\:$87}&\\
 102& 14.33  & 31.4   &  9.2 & WC9?d & {\changed ~$\:$97}&{\bf{Q~9}}, {\changed MGM~5-9}, WR~102db, qF~258, GCS~3-3, narrow and weak lines\\
 103& 14.20  & 31.4   & 12.7 & O7-9 I f& {\changed ~$\:$97}:&\\
 104& 16.34  & 31.1   & 13.6 & O9-B2 I-II f?& {\changed ~$\:$97}&\\
 105& 15.60  & 31.2   & 12.6 & O7-9 I f? & {\changed 107}&\\
 106& 16.31  & 30.4   & 12.7 & O7-9.5 I & {\changed 112}&\\
 107& 16.05  & 30.1   & 13.9 & O7-9 I f?& {\changed 106}:&\\
 108& 15.46  & 29.6   & 10.7 & M3 I & {\changed 127}&qF~269, [GMC99] D~322, [NWS90]~A\\
 109& 14.99  & 29.7   & 13.0 & O7-9 I f?& {\changed ~$\:$97}:&\\
 110& 15.09  & 29.4   & 10.6 & O6-8 I f (Of/WN?)& {\changed 117}& {\bf{Q~15}}, qF~270S, MGM~5-15, [LFG99] QR~4, {\changed [NWS90] C}\\
\s
 111& 14.99  & 29.1   & 12.4 & O8-9 I f?& {\changed 107}:&\\
 112& 16.72  & 28.4   & 12.8 & K4 I-II & {\changed $\:$~$\:$~6}f&\\
 113& 16.07  & 28.2   & 12.5 & K2 I & {\changed 136}&\\
 114& 14.19  & 28.2   & 13.9 & K4 II & {\changed 137}&\\
 115& 15.16  & 27.8   &  8.6 & M2 I & {\changed 137}& {\bf{Q~5}}, qF~270N, V~4646
  Sgr, {\changed MGM~5-5}\\
 116& 14.76  & 27.9   & 12.2 & M1 I-II & {\changed ~$\:$57f}&\\
 117& 14.91  & 27.8   & 12.1 & K0? I & {\changed 142}:&\\
 118& 16.05  & 27.4   & 11.5 & O6-9 I f?& {\changed ~$\:$56f}&\\
 119& 14.80  & 27.5   & 12.7 & K0? I & {\changed ~$\:$17}:f&\\
 120& 16.59  & 27.2   & 13.9 & O5-9 I-II f?& {\changed 106}:&\\
\s
 121& 14.36  & 27.1   & 13.4 & K1 I & {\changed ~$\:$97}&\\
 122& 14.25  & 26.9   & 11.6 & O7-9.7 I e& {\changed 107}&\\
 123& 15.72  & 26.7   & 12.8 & K1 I & {\changed ~$\:$36}f&\\
 124& 15.66  & 26.8   & 13.5 & K0? I & {\changed ~$\:$32}:f&\\
 125& 14.50  & 26.6   & 13.5 & O7-9 I-II & {\changed ~$\:$97}:&\\
 126& 16.56  & 26.2   & 12.5 & O7-9 I f?& {\changed 106}:&\\
 127& 14.13  & 26.1   & 13.5 & K1 II & {\changed ~$\:$97}:&\\
 128& 14.62  & 25.9   & 13.5 & O3-5 I & {\changed ~$\:$97}:&\\
 129& 14.77  & 25.3   & 13.4 & K3 II & {\changed 137}&\\
 130& 16.44  & 24.9   & 13.9 & O7.5-9 I-II e?& {\changed ~$\:$96}:&\\
\s
 131& 14.22  & 24.9   & 13.5 & K2 II & {\changed ~$\:$17}f&\\
 132& 14.43  & 24.7   & 13.0 & O7-9 I-II f?& {\changed ~$\:$97}&\\
 133& 16.02  & 24.3   & 13.4 & K3 II & {\changed 156}:&\\
 134& 15.28  & 24.4   & 13.5 & O9-B2 I f?& {\changed ~$\:$97}:&\\
 135& 15.05  & 24.4   & 11.7 & K4 I & {\changed \,-33}f&\\
\s
 136& 16.51  & 23.9   & 11.2 & M2 I & {\changed ~$\:$76}&\\
 137& 16.60  & 23.6   & 13.1 & K2 I & {\changed 176}&\\
 138& 16.30  & 22.9   & 11.9 & M1 I & {\changed 146}&\\
 139& 16.61  & 22.8   & 13.5 & O7-9 I f?& {\changed 101}:&\\
 140& 15.41  & 22.8   & 14.2 & K3 II & {\changed 133}:&\\
\s
 141& 14.99  & 22.3   & 11.5 & O9.7-B1 I & {\changed 103}:&\\
 142& 14.30  & 22.1   & 13.5 & K3 II & {\changed 122}&\\
 143& 16.02  & 21.4   & 10.5 & O7-B0 I & {\changed 115}:& qF~301, [GMC99] D~271\\ 
 144& 15.34  & 20.4   & 11.8 & O7-9 I & {\changed 124}:&\\
 145& 16.01  & 20.3   & 12.7 & O7-9 I & {\changed 115}:&\\
 146& 15.49  & 20.1   &  8.7 & O6-8 I f?& {\changed 124}& {\bf{Q~14}}, qF~307A, MGM~5-14\\
 147& 14.27  & 19.96  & 12.8 & M0 II & {\changed 126}&\\
 148& 15.05  & 18.9   & 11.3 & O7-9 I & {\changed 109}&\\
 149& 14.00  & 18.6   & 12.1 & O7-9.7 I & {\changed 101}&\\
 150& 15.61  & 18.1   & 13.4 & K1 I & {\changed 169}:&\\
\s
 151& 14.14  & 17.99  & 14.0 & K0? II & {\changed ~$\:$56f}&\\
 152& 16.08  & 17.6   & 14.0 & O6.5-8 I-II f& {\changed ~$\:$95}&\\
 153& 15.18  & 17.5   & 13.5 & K4 II & {\changed ~$\:$59f}&\\
 154& 14.85  & 17.3   & 12.5 & O7-9 I & {\changed 109}:&\\
 155& 15.36  & 17.1   & 13.9 & K2 II & {\changed ~$\:$69}:&\\
 156& 15.61  & 16.9   & 10.4 & M0 I & {\changed 169}& [GMC99] D~307\\
 157& 16.06  & 16.9   & 14.5 & O3-4 II/III? & {\changed 115}&\\
 158& 14.05  & 16.6   & 10.5 & WN9 & {\changed 106}& WR~102d, q{\changed F}~320, [LGF99] QR~8, {\changed [GMC99] D278?}\\
 159& 15.94  & 16.4   & 13.9 & O7-9 I& {\changed 105}&\\
 160& 15.90  & 15.8   & 14.6 & K1 II & {\changed 165}&\\
\hline		
\hline     		     	        
\end{longtable}
 \begin{list}{}{}
  \item{{\it{List of catalogs (SIMBAD identifier and reference):}}}
  \item{GCS -- \citet{GCS83} \quad \quad Q $\rightarrow$ GMM {\changed
  catalog} --
  \citet{GMM90} \quad \quad [NWS90] -- \citet{NWS90}}
  \item{MGM -- {\changed \citet{MGM92, MGM1994}} \quad \quad [NHS93] --
  \citet{Nagata+1993} \quad \quad qF -- \citet{FMM99}}
  \item{[FMG99] -- \citet{FMG99} \quad \quad [GMC99] -- \citet{GMC99}
  \quad \quad [LFG99] -- \citet[2003, 2005]{Lang+1999}}
  \item{[GSL2002] -- \citet{GSL2002} \quad \quad [LY2004] --
  \citet{LY2004} \quad \quad WR -- \citet{vdH2006}}
  \item{}
  \item{{\it{Spectral classification:}}}
  \item{OB stars -- \citet{Hanson+1996, Hanson+2005} and \citet{Morris+1996}:
       using \ion{H}{i} 2.1661\,$\mu$m {\changed Br}\,$\gamma$, \ion{He}{i}
       2.0587\,$\mu$m, 2.1127/37~$\mu$m, 2.1499\,$\mu$m,
       2.1623\,$\mu$m, \ion{He}{ii} 2.1891\,$\mu$m, \ion{C}{iv}
       triplet around 2.0796\,$\mu$m, \ion{N}{iii}/\ion{C}{iii} 2.1155\,$\mu$m}
  \item{WR stars -- \citet{Crowther+2006}:
       applying line ratio \ion{C}{iv}\,(2.079)/\ion{C}{iii}\,(2.108)
       for WC and \ion{He}{ii}\,(2.189)/Br\,$\gamma$\,(2.166) for WN stars}
  \item{KM giant stars -- \citet{Wallace-Hinkle1997,
       Kleinmann-Hall1986, Goorvitch1994}: $^{12}$CO and $^{13}$CO bands,
       subclasses following \cite{Gonzalez-Fernandez+2008}, see
       text Sect.\,\ref{subsec:spec-class}}
  \item{}
  \item{{\it{Magnitudes:}}}
  \item{Synthetic $K_\textrm s$ magnitudes from our calibrated
  spectra, see Sect.\,\ref{subsec:photometry} for details.}
  \item{}
  \item{{\it{Radial velocity and Cluster membership:}}}
  \item{$RV$ measured with the position of the Br~$\gamma$ line for
  {\changed early-type} stars and CO (2-0) band head for {\changed late-type}
  stars. Uncertain measurements are marked with ``:'', foreground
  objects are indicated by ``f'', see Sect.~\ref{subsec:clustermembers}.}  
 \end{list}
\end{landscape}
}
}}


\Online
\label{onlinematerial}

The Online Material presents the $K$-band spectra (1.95 -- 2.45\,$\mu$m)
of 160 stars in the field of the Quintuplet cluster. 98 stars are of early
spectral type (O, B and WR), and 62 of late type (K and M). A few spectra
are dominated by dust emission.

The ``LHO'' numbering of the stars refers to Table\,\ref{tab:catalog}
of the main paper. 
The spectra are flux-calibrated, and binned in wavelength over 4\,\AA. The
wavelength scale refers to the local standard of rest.

\end{document}